\pdfoutput=1
\RequirePackage{ifpdf}
\ifpdf 
\documentclass[pdftex]{sigma}
\else
\documentclass{sigma}
\fi

\numberwithin{equation}{section}

\begin{document}

\allowdisplaybreaks

\renewcommand{\thefootnote}{$\star$}

\newcommand{\arXivNumber}{1603.07278}

\renewcommand{\PaperNumber}{069}

\FirstPageHeading

\ShortArticleName{Random Tensors and Quantum Gravity}

\ArticleName{Random Tensors and Quantum Gravity\footnote{This paper is a~contribution to the Special Issue on Tensor Models, Formalism and Applications. The full collection is available at \href{http://www.emis.de/journals/SIGMA/Tensor_Models.html}{http://www.emis.de/journals/SIGMA/Tensor\_{}Models.html}}}

\Author{Vincent RIVASSEAU}

\AuthorNameForHeading{V.~Rivasseau}

\Address{Laboratoire de Physique Th\'eorique, CNRS UMR 8627, Universit\'e Paris XI, \\
F-91405 Orsay Cedex, France}
\Email{\href{mailto:rivass@th.u-psud.fr}{rivass@th.u-psud.fr}}
\URLaddress{\url{http://www.rivasseau.com/}}

\ArticleDates{Received March 23, 2016, in f\/inal form July 06, 2016; Published online July 15, 2016}

\Abstract{We provide an informal introduction to tensor f\/ield theories and to their associated renormalization group. We focus more on the general motivations coming from quantum gravity than on the technical details. In particular we discuss how asymptotic freedom of such tensor f\/ield theories gives a concrete example of a natural ``quantum relativity'' postulate: physics in the deep ultraviolet regime becomes asymptotically more and more independent of any particular choice of Hilbert basis in the space of states of the universe.}


\Classification{60B20; 81T15; 81T16; 81T17; 82B28}

\renewcommand{\thefootnote}{\arabic{footnote}}
\setcounter{footnote}{0}

\section{Introduction}

\emph{In physics, a field is a physical quantity that has a value for each point in space and time.}

This current wikipedia def\/inition of a f\/ield reminds us how much a~preexistent classical space-time with its associated notion of locality is deeply ingrained in our common representation of the physical world. However there is an ``emerging'' consensus than an \emph{ab initio} theory of quantum gravity requires to at least modify and probably abandon altogether the concepts of absolute locality and absolute classical space-time \cite{Oriti:2006ar,Seiberg:2006wf,Sindoni:2011ej}. It is a task for many generations of theoretical physicists to come, requiring even more work than when classical physics had to be abandoned in favor of quantum mechanics.

The tensor track \cite{Rivasseau:2011hm,Rivasseau:2012yp,Rivasseau:2013uca} can be broadly described as a step in this direction and more precisley as a program to explore (Euclidean) quantum gravity as a (Euclidean) quantum f\/ield theory \emph{of} space-time rather than \emph{on} space-time, relying on a specif\/ic mathematical formalism, namely the modern theory of random tensors \cite{Bonzom:2012hw, Gurau:2009tw,Gurau:2011kk,Gurau:2011xp} and their associated $1/N$ expansions \cite{Bonzom:2012wa,Bonzom:2015axa, Gurau:2010ba,Gurau:2011xq,Gurau:2011aq}, which generalizes in a natural way the theory of random vectors and random matrices~\cite{DiFrancesco:1993cyw}. In this formalism observables, interactions, and Feynman graphs are all represented by regular edge-colored graphs.

It is both simple and natural but also general enough to perform quantum sums over space-times in any dimension, pondering them with a discrete analog of the Einstein--Hilbert action \cite{ambjorn}. In three dimensions the formalism sums over all topological manifolds, and in four dimensions over all triangulated manifolds\footnote{Some topological manifolds in four dimensions, such as the $E_8$ manifold, cannot be triangulated as a simplicial complex but it is not clear they are relevant for quantum gravity.} with all their dif\/ferent smooth structures \cite{crystal2,crystal3,crystal4,crystal5,crystal1, scorpan}.

More precisely the tensor track proposes to explore renormalization group f\/lows \cite{Geloun:2015qfa,Benedetti:2014qsa,Benedetti:2015yaa,Eichhorn:2013isa,Krajewski:2015clk} in the associated tensor theory space~\cite{Rivasseau:2014ima}. In doing this, it can retain several of the most characteristic aspects of the successful quantum f\/ield theories of the standard model of particle physics, such as perturbative renormalizability \cite{Geloun:2013saa,BenGeloun:2011rc,Geloun:2012fq,Carrozza:2013wda,Carrozza:2012uv,Samary:2012bw} and asymptotic freedom \cite{BenGeloun:2012yk,Geloun:2012qn,BenGeloun:2012pu,Rivasseau:2015ova,Samary:2013xla}. Also at least in the simplest cases, random tensor models and f\/ield theories can be contructed non-perturbatively \cite{Delepouve:2014bma,Delepouve:2014hfa,Gurau:2013pca,Gurau:2013oqa,Lahoche:2015yya,Lahoche:2015zya}.

The tensor track can be considered both as a generalization of random matrix models, successfully used to quantize two-dimensional gravity~\cite{DiFrancesco:1993cyw}, and as an improvement of group f\/ield theory \cite{Baratin:2011hp,Geloun:2009pe,boulatov,laurentgft,Krajewski:2012aw,Oriti:2007qd,Oriti:2007vf} motivated by the desire to make it renormalizable \cite{Geloun:2011cy,Geloun:2010nw,FreiGurOriti, Rivasseau:2011xg}. It proposes another angle for the study of dynamical triangulations \cite{ambjorn-book,Ambjorn:2013apa,scratch}. It also relies on non-commutative f\/ield theory and in particular on the Grosse--Wulkenhaar model \cite{Disertori:2006nq,Disertori:2006uy,Grosse:2004by,Grosse:2004yu,Grosse:2009pa,Grosse:2012uv,Grosse:2014lxa,Grosse:2015fka} as a strong source of inspiration.

This brief paper is intended as an introduction in non-technical terms to this approach. We warn the reader that it overlaps strongly with previous
reviews and papers of the author such as \cite{Rivasseau:2011hm,Rivasseau:2012yp,Rivasseau:2014ima,Rivasseau:2013uca,Rivasseau:2015ova}.

\section{Historical perspective}

After a summer course with J.~Schwinger and a long road trip through the United States with R.~Feynman, F.~Dyson understood how to relate Feynman's functional integration (sum over histories) to Schwinger's equations~\cite{dyson}. The iterative solution of these ``Schwinger--Dyson'' equations is then indexed by Feynman's graphs. Quantum f\/ield theory was born.

From the start it had to struggle with a famous problem: the amplitudes associated to Feynman graphs contain ultraviolet divergencies. Any quick f\/ix of this problem through imposing an ultra-violet cutof\/f violates some of the most desired properties of the theory such as Euclidean invariance or Osterwalder--Schrader positivity, the technical property which allows continuation to real time and unitarity, hence ultimately ensures that quantum probabilities add up to~1, as they should in any actual experiment.

The full solution of the problem, namely renormalization, took some time to polish. The textbook example is the one of an interacting theory obtained by perturbing a non-local free theory by a local interaction with a small coupling constant. The free theory is represented by a Gaussian measure based on a non-local propagator, such as $1/(p^2+m^2)$, which becomes however of shorter and shorter range, hence is \emph{asymptotically local} in the ultraviolet limit. In this case the key ingredients allowing renormalization to work are the notion of scale and the locality principle: high scale (ultraviolet) connected observables seem more and more local when observed with lower scale (infrared) propagators. It also requires a power counting tool which allows to classify observables into relevant, marginal and irrelevant ones. Under all these conditions, the ultraviolet part of divergent Feynman graphs can be absorbed and understood as modifying the ef\/fective value of the coupling constant according to the observation scale. This is the standard example of a renormalization group f\/low.

The renormalization group of K.~Wilson and followers is a powerful generalization of the previous example. The action takes place in a certain \emph{theory space} which is structured by the locality principle: the local potential approximation corresponds to the purely local powers of the f\/ield at the same point. Such local operators can be dressed with an arbitrary number of derivatives, creating a hierarchy of ``quasi-local'' operators. The operator product expansion expands the average value of any non-local operator, when observed through infrared probes, into a~dominant local part plus correction terms with more and more derivatives, hence it is an expansion into quasi-local operators which generalizes the multipolar expansion of close conf\/igurations of charges in classical electrostatics. The (irreversible) renormalization group f\/lows from bare (ultraviolet) to ef\/fective (infrared) actions by integrating out (quasi-local) f\/luctuations modes. Fixed points of the f\/low need no longer be Gaussian, but interesting ones should still have only a small number of associated relevant and marginal directions. Indeed it is under this condition that the corresponding physics can be described in terms of only a few physical parameters.

Scales and locality play the fundamental role in this standard picture of a renormalization group analysis. Locality is also at the core of the mathematically rigorous formulation of quantum f\/ield theory. It is a key Wightman axiom \cite{SW} and in algebraic quantum f\/ield theory~\cite{haag} the fundamental structures are the algebras of \emph{local observables}. It is therefore not surprising that to generalize quantum f\/ield theory and the locality principle to an abstract framework independent of any preexisting space-time takes some time.

Such a development seems however required for a full-f\/ledged \emph{ab initio} theory of quantum gravity. Near the Planck scale, space-time should f\/luctuate so violently that the ordinary notion of locality may no longer be the relevant concept. Among the many arguments one can list pointing into this direction are the Doplicher--Fredenhagen--Roberts remark that to distinguish two objects closer than the Planck scale would require to concentrate so much energy in such a~little volume that it would create a~black hole, preventing the observation~\cite{DFR}. String theory, which (in the case of closed strings) contains a gravitational sector, is another powerful reason to abandon strict locality. Indeed strings are one-dimensional \emph{extended} objects, whose interaction cannot be localized at any space time point. Moreover, closed strings moving in compactif\/ied background may not distinguish between small and large such backgrounds because of dualities that exchange their translational and ``wrapping around'' degrees of freedom. Another important remark is that two and three-dimensional pure quantum gravity are topological theories. In such theories observables, being functions of the topology only, cannot be localized in a particular region of space-time.

Remark however that four-dimensional gravity is certainly not a purely topological theory, as conf\/irmed by the recent observation of gravitational waves. It is nevertheless an educated guess that pure (Euclidean) four-dimensional quantum gravity should not completely loose any topological character. In other words modif\/ications of the topology of space-time should \emph{not} be strictly forbidden. Indeed space-times of dif\/ferent (complicated) topologies can interpolate between f\/ixed boundaries, even when the latter have simple topology. Suppressing some of them because they do not have a simple topology would be similar to suppressing arbitrarily instantons from Feynman's sum over histories, and this is known to lead to quantum probabilities no longer adding up to~1.

Space-time dimension four is the f\/irst in which gravity has local degrees of freedom but it is also the f\/irst dimension in which exotic smooth structures can appear on a f\/ixed topological background manifold~\cite{scorpan}. It would be desirable to establish a strong link between these two properties of dimension four, in particular because smooth structures in four dimensions are intimately related to gauge theories~\cite{Donaldson}, which describe the other physical interactions, electroweak and strong, in the standard model. Staying within the scope of this small review, let us simply remark again that the basic building blocks proposed by the tensor track, namely tensor invariants, are dual to piecewise linear manifolds with boundaries. Therefore they precisely distinguish in four dimension all smooth structures associated to a given topological structure~\cite{crystal3,crystal4,crystal5}.

\section{The quantum relativity principle} \label{qrp}
Let us try to forget for a minute the enormous amount of technical work spent on various theories of quantum gravity, and consider with a fresh eye a very general, even naive question: what are the most basic mathematical tools and physical principles at work in general relativity and quantum mechanics? How could we most naturally join them together?

The very name of general relativity suggests how Einstein derived it: as a consequence of the independence of the laws of physics under any particular choice of space-time coordinates. In a sense our coordinates systems are man-made: the Earth does not come equipped with meridians or parallels drawn on the ground. So it is natural to expect the general laws of physics not to depend on such man-made coordinates. The corresponding symmetry group is the group of dif\/feomorphisms of space-time. But since this group depends on the underlying space-time manifold, in particular of its topology, the classical general relativity principle is not fully background invariant.

\looseness=-1
In quantum mechanics, instead of space-time trajectories, the basic objects are states, ele\-ments of a Hilbert space and the observables are operators acting on them. An important observation is that although there are many dif\/ferent manifolds in a given dimension, each with its own dif\/ferent dif\/feomorphism group, there is only a \emph{single} Hilbert space of any given dimension $N$, denoted ${\mathcal H}_N$ (up to isomorphism, which here means up to change of orthonormal basis).

There is also a single \emph{separable} inf\/inite-dimensional Hilbert space, namely \emph{the} Hilbert space~${\mathcal H}$. Separable means here that~${\mathcal H}$ admits a countable orthonormal basis. ${\mathcal H}$ can be identif\/ied with the space of square integrable series $\ell_2 ( \mathbb{N} )$, or with $\ell_2 ( \mathbb{Z}) = L^2 ( {\rm U}(1))$ or with the $L^2$ space of square integrable functions on \emph{any} Riemannian manifold of \emph{any} f\/inite dimension. It is therefore a truly background-independent mathematical structure.

Our universe certainly contains a huge number of degrees of freedom, both of geometric (gravitational waves) and matter type. Hence ${\mathcal H}= \lim\limits_{N \to \infty} {\mathcal H}_N$ seems the right mathematical starting point for a background independent quantum theory of gravity. That this was not emphasized more in the early days of quantum mechanics probably only means that absolute space-time was still a deeply ingrained notion in the minds of the theoretical physicists at that time.

It is then tempting to extend the general relativity principle into a quantum relativity principle by postulating that the laws of physics should be invariant under the symmetry group of \emph{that unique} Hilbert space~${\mathcal H}$, namely independent of any preferred choice of an orthonormal basis in~${\mathcal H}$. Just like for coordinates systems, an orthonormal basis seems to be observer-made.

In a very large universe with $N$ degrees of freedom it would amount to postulate the ${\rm U}(N)$ invariance of physics. However such a postulate quickly appears a bit too extreme. The only (polynomial) ${\rm U}(N)$ invariants in a f\/inite-dimensional Hilbert space ${\mathcal H}_N$ are polynomials in the scalar product. We know that quantum states, being rays in Hilbert space, can be normalized to~1, hence there is no fully ${\rm U}(N)$-invariant physical observables which can distinguish between two states hence interesting observables need to break this full symmetry.

Beauty has sometimes been def\/ined as a ``slightly broken symmetry'' and this may also be a~good def\/inition for physics. Just like ordinary quantum f\/ield theory is not exactly local but only asymptotically local in the ultraviolet regime, the giant~${\rm U}(N)$ invariance of~${\mathcal H}_N$ at very large~$N$ could be asymptotic in a~certain regime which by analogy we should still call the ultraviolet regime. In practice it has to be broken in any actual experiment. If for no other reason, it should be at least because we are f\/inite size observers in an ef\/fective geometry of space-time. Hence we are led to consider an attenuated version of invariance of physics under change of basis, namely

{\bf Quantum relativity principle.} \emph{In the extreme ``ultraviolet'' regime $N \to \infty$ the laws of physics should become \emph{asymptotically} independent of any preferred choice of basis in the quantum Hilbert space describing the universe.}

Of course we have to explain in more detail the meaning of the words ``asymptotically independent in the extreme ultraviolet regime $N \to \infty$''. It has to be understood in a renormalization group sense. To def\/ine an abstract (space-time independent) renormalization group and its corresponding asymptotic ultraviolet regime is the main goal of the tensor track, and it requires several ingredients. We need f\/irst an initial device to break the ${\rm U}(N)$-symmetry group and allow to label and regroup the degrees of freedom in a way suited for a~renormalization group analysis. In particular it should allow to group together the degrees of freedom into renorma\-li\-za\-tion group \emph{slices}. The ones with many degrees of freedom will be called the ultraviolet slices, and the ones with less degrees of freedom the infrared slices. In tensor f\/ield theories this device is a ${\rm U}(N)$-breaking propagator for the free theory, whose spectrum allows to label and regroup the degrees of freedom of~${\mathcal H}_N$, exactly like Laplacian or Dirac-based propagators do in ordinary quantum f\/ield theory\footnote{Even if one does not like the idea of a non-invariant propagator, to set up a renormalization group analysis can still be done by breaking ${\rm U}(N)$ invariance at the level of the cutof\/fs which separate degrees of freedom into f\/luctuations and background.}.

Once such a device is in place, the renormalization group always then means a decimation, to f\/ind the ef\/fective infrared theory after integrating ultraviolet slices of the theory. Asymptotic invariance in the ultraviolet regime means that the bare action should be closer and closer to an exactly ${\rm U}(N)$-invariant action. If we agree on the ``quantum relativity principle'' above and on this general strategy, the next step is to search for the most natural symmetry breaking pattern that could occur on the path from the extreme ultraviolet regime to ef\/fective symmetry-broken infrared actions.

\section{Random tensors}

\looseness=-1
We know that the Hilbert space of a composed physical system is not the direct sum but the tensor product of the Hilbert spaces of its constituents, and this aspect is critical to entanglement, a~basic feature of quantum mechanics. Hence the most natural pattern for symmetry breaking of ${\rm U}(N)$ invariance is to change vectors into tensors of a~certain rank~$d$. This breaks the ${\rm U}(N)$ group to a smaller group. More precisely if $N$ factors as a product of two integers\footnote{This is generically possible at large $N$ because of the rarefaction of prime numbers.} $N=N_1 . N_2 $, we can break ${\rm U}(N)$ into ${\rm U}(N_1) \otimes {\rm U}(N_2)$, the natural symmetry group of random rectangular~$N_1$ by~$N_2$ Wishart matrices~\cite{Wishart} in the tensor space ${\mathcal H}_{N_1} \otimes {\mathcal H}_{N_2}$. Such Wishart matrices are not just rectangular arrays of random numbers. Their polynomial interactions and observables, which are product of traces, have the reduced ${\rm U}(N_1)\otimes {\rm U}(N_2)$ symmetry rather than a full ${\rm U}(N_1 N_2)$ vector symmetry.

Continuing along this line, if $N=N_1 . N_2 \dots N_d$, we can break ${\rm U}(N)$ into ${\rm U}(N_1) \otimes \cdots \otimes {\rm U}(N_d)$, the natural symmetry group of tensors of rank $d$ in a tensor space ${\mathcal H}_{N_1} \otimes \cdots \otimes{\mathcal H}_{N_d}$. The reduced symmetry corresponds to invariance under independent change of coordinates in each of the tensor factor, namely in each space ${\mathcal H}_{N_1}, \dots, {\mathcal H}_{N_d}$. The invariant connected polynomials associated to this symmetry are exactly connected bipartite $d$-regular edge-colored graphs. Such connected bipartite $d$-regular edge-colored graphs on $n$ vertices can be enumerated~\cite{Geloun:2013kta} and their number~$Z_d(n)$, for $d \ge 3$ grows rapidly with~$n$:
\begin{alignat*}{3}
& Z_1 (n) = 1, 0, 0, 0, 0, \dots, \qquad && Z_3(n) = 1, 3, 7, 26, 97, 624, \dots,&\\
& Z_2 (n) = 1, 1, 1, 1, 1, 1, 1, \dots,\qquad && Z_4(n) = 1, 7, 41, 604, 13753, \dots.&
\end{alignat*}

Tensor models with canonical trivial propagator and polynomial invariant interactions have a perturbative expansion indexed by Feynman graphs which, under expansion of the internal structure of their bipartite $d$-regular edge-colored vertices, become bipartite $(d + 1)$-regular edge-colored graphs. They typically admit an associated 1/N expansion whose leading graphs, called melons~\cite{Bonzom:2011zz}, are in one-to one correspondence with $(d + 1)$-ary trees\footnote{This is true when the interactions of these models are themselves melonic; for other cases, see \cite{Lionni}.}.

This observation has deep geometric consequences. These graphs precisely encode all piecewise linear orientable\footnote{Non-orientable manifolds can be included by considering only real $O(N)$ invariance, hence loosing bipartiteness but not the \emph{colors}.} manifolds (with boundaries) in dimension $d$ \cite{crystal1}. We could therefore consider the reduced symmetry ${\rm U}(N_1) \otimes \cdots \otimes {\rm U}(N_d)$ as a quantum precursor of the coordinate invariance of general relativity and the invariant connected polynomials associated to this symmetry as precursors of observables. More precisely, both interactions and observables of the theory can be associated with (colored triangulations of) $d$-dimensional spaces and the Feynman graphs of the theory are associated with (colored triangulations of) $(d+1)$-dimensional space-time \cite{Bonzom:2012hw}.

The single \cite{Bonzom:2011zz} and double \cite{Dartois:2013sra,GurauSchaeffer,Gurau:2015tua} scaling limits of such tensor models for rank $d\ge 3$ have now been identif\/ied and correspond to continuous random trees~\cite{Gurau:2013cbh}. This seems at f\/irst sight a step backwards. Indeed the $1/N$ expansion of vector models is also dominated by trees, whereas the $1/N$ expansion of matrix models is dominated by the more complicated planar maps. Branched polymers, like planar maps are certainly not a good approximation to our current macroscopic universe, but planar maps seem closer, since they have higher Hausdorf\/f and spectral dimension. However tensor invariant interactions and the sub-leading structure of the $1/N$ tensorial expansion are much richer than their vector and matrix counterparts. This is not too surprising, given that tensor models of rank $d$ can ef\/fectuate a statistical sum over all manifolds (and many pseudo-manifolds) in dimension $d$.

To go beyond single and double scaling of the simplest models and to f\/ind more interesting infrared limits and phase transitions probably will require to combine analytic and numerical methods. Tensor group f\/ield theory (TGFT), which we now describe, adds a Laplacian to the propagator of the tensor models to equip them with a full-f\/ledged notion of renormalization group. It generalizes in a natural way the matrix renormalization group of the Grosse--Wulkenhaar model~\cite{Disertori:2006nq,Disertori:2006uy,Grosse:2004by,Grosse:2004yu,Grosse:2009pa,Grosse:2012uv,Grosse:2014lxa,Grosse:2015fka}. Such tensor theories become then suited for a functional renormalization group analysis \cite{Geloun:2015qfa,Benedetti:2014qsa,Benedetti:2015yaa,Eichhorn:2013isa,Krajewski:2015clk}.

Asymptotic ${\rm U}(N^d)$ invariance and the ``quantum relativity principle'' will then be recovered in the deep ultraviolet regime if the theory is \emph{asymptotically free}. This is due to the combination of two facts. First the interaction, which broke ${\rm U}(N^d)$ to ${\rm U}(N)^{\otimes d}$ asymptotically vanishes in the ultraviolet regime because of this asymptotic invariance. Second the propagator itself, because it is of the inverse Laplacian type, has a smaller and smaller relative variation in the ultraviolet regime between dif\/ferent frequencies:
\begin{gather*}
 \frac{\delta}{\delta p} \big(p^2 + m^2\big)^{-1} \simeq \frac{p}{(p^2 + m^2)^2} \ll_{p \to \infty} \big(p^2 + m^2\big)^{-1}.
\end{gather*}
Full ${\rm U}(N^d)$ asymptotic invariance is then in fact recovered in the ultraviolet limit precisely in the same asymptotic sense than locality is asymptotically recovered in the ultraviolet regime of ordinary (Euclidean) quantum f\/ield theory: high energy subgraphs look almost local when observed through infrared probes with low resolution. We could also say that in such a scenario the degrees of freedom of the universe behave more and more as the molecules of a perfect gas in the extreme ultraviolet, high-temperature regime.

\section{Tensor group f\/ield theories}

As already explained, the key dif\/ference between random matrix models and non commutative f\/ield theory on Moyal space lies in the modif\/ication of the propagator. Breaking the ${\rm U}(N)$ invariance of matrix models at the propagator level can be done in many ways, but the simplest way was found in the Grosse--Wulkenhaar (GW) model~\cite{Grosse:2004yu} on ${\mathbb R}^4$ equipped with the Moyal product. It is a matrix model with quartic coupling and a propagator which can be interpreted as the sum of the Laplacian and an harmonic potential on~${\mathbb R}^4$. It is very natural in the matrix base which expresses the Moyal star product as a matrix product. Such a propagator has a non-trivial spectrum, hence breaks the ${\rm U}(N)$ invariance of the theory. It allows to distinguish the infrared (small values of the base index, large eigenvalues of the propagator) from the ultraviolet regime (large values of the base index, small eigenvalues of the propagator). There is an associated renormalization group f\/low between these two regimes.

Power counting in such models is entirely governed by the underlying $1/N$ expansion (divergent graphs are the planar~2 and~4 point graphs with a single external face), hence the corresponding matrix renormalization group~\cite{Eichhorn:2013isa} can be considered a kind of continuous version of the $1/N$ expansion. It is remarkable that this model is asymptotically scale invariant, that is in the ultraviolet regime the beta function tends to zero~\cite{Disertori:2006nq,Disertori:2006uy, Grosse:2004by}. Its planar sector has also recently been beautifully solved~\cite{Grosse:2009pa,Grosse:2012uv}, showing that four-dimensional integrability, like four-dimensional asymptotic safety, is not limited to the famous example of supersymmetric Yang--Mills $N=4$ QFT but is more general and does not \emph{require} supersymmetry. Also in a~rather surprising way the model at negative coupling very probably obeys all Wightman axioms except clustering~\cite{Grosse:2014lxa,Grosse:2015fka}.

In a completely analogous manner, one can def\/ine TGFT's with tensorial interactions and a~soft breaking of the tensorial invariance of their propagator. Renormalization is again a~continuous version of the $1/N$ expansion and the divergent graphs reduce to the melonic sector instead of the planar sector. The group structure allows to interpret the f\/ields either as rank $d$ tensors or as ordinary f\/ields def\/ined on $G^d$ where $G$ is the Lie group. The propagator is the inverse of the sum of the Laplacians on each factor in $G^d$. In the simplest case, namely $G={\rm U}(1)$, one can identify at any rank which are the super-renormalizable and just renormalizable interactions. Surprisingly perhaps, at rank/dimension~4 the just renormalizable interactions are of order~6, not~4 \cite{BenGeloun:2011rc,Geloun:2012fq,BenGeloun:2012pu}. The quartically interacting model is just renormalizable at rank/dimension~5. It is also just renormalizable at rank/dimension 3 if the propagator has Dirac-type rather than Laplace-type power counting. It is remarkable that the corresponding tensorial RG f\/low displays the generic property of asymptotic freedom, at least for quartic interactions~\cite{BenGeloun:2012yk,Geloun:2012qn, BenGeloun:2012pu}; see however~\cite{Carrozza:2014rba} for the subtle issue of interactions of order~6.

\looseness=-1
The bridge between matrix and tensor renormalizable models has been further reduced in~\cite{Geloun:2012bz}, in which new families of renormalizable models are identif\/ied. It is shown that the rank~3 tensor model def\/ined in~\cite{BenGeloun:2012pu}, and the Grosse--Wulkenhaar models in two and four dimensions generate three dif\/ferent classes of renormalizable models by modifying the power of the propagator. A~review on this class of models extends and generalizes this analysis~ \cite{Geloun:2013saa}. It classif\/ies the models of matrix or tensor type with a propagator which is the inverse of a sum of momenta of the form~$p^{2a}$, $a \in ]0,1]$. Inf\/inite towers of (super- and just-) renormalizable matrix models are found.

The emerging picture is a simple classif\/ication of quite a large class of Bosonic renorma\-li\-zab\-le non-local QFTs of the matrix or tensor type. The theories of scalar and vector type are neither asymptotically free nor safe since they have no wave function renormalization at one loop. The theories of matrix type are generically asymptotically safe (specially for instance all~$\operatorname{Tr}[M^4]$ models), since the wave function renormalization exactly compensates the coupling constant renormalization in the ultraviolet regime. The theories of tensor type are generically asymptotically free since the wave function renormalization wins over the coupling constant renormalization. As discussed in Section~\ref{asymfree}, which we essentially reproduce from~\cite{Rivasseau:2015ova} for self-content, this is a robust fact since it is rooted in the combinatorial structure of the quartic vertex, which is dif\/ferent for vector, matrix and tensor models.

The combinatorics of renormalization is neatly encoded in a Connes--Kreimer combinatorial Hopf algebra. See~\cite{Krajewski:2012is} for a version of this algebra adapted to multi-scale analysis. This algebra has been explicited in~\cite{Raasakka:2013kaa} for the case of the model def\/ined in~\cite{BenGeloun:2011rc,Geloun:2012fq}. It dif\/fers signif\/icantly from previous Connes--Kreimer algebras, due to the involved combinatorial and topological properties of the tensorial Feynman graphs.

\subsection{Laplacian with gauge projector}

TGFTs which include a ``Boulatov-type'' projector in their propagator \cite{boulatov} deserve a category of their own as they are both more dif\/f\/icult to renormalize and closer to the desired geometry in the continuum. Their characteristic feature is to include an average over a new Lie group variable acting simultaneously over all tensor threads in the propagator. This average implements in three dimensions the constraints of the $BF$ theory, which is classically equivalent to three-dimensional general relativity. In four dimensions one may similarly use projectors in order to implement the Plebanski simplicity constraints of four-dimensional gravity~\cite{Rovelli}. In a~Feynman graph, once the vertex variables have been integrated out, the Feynman amplitude appears as an integral over one variable for each edge of a product of delta functions for each face. Each such delta function f\/ixes to the identity the ordered product of the edge variables around the face. Therefore it can be interpreted as a theory of simplicial geometry supplemented with a~discrete gauge connection at the level of the Feynman amplitudes of the theory which has trivial holonomy around each face of the Feynman graph, hence around each $d-2$ cell of the dual triangulation.

To introduce renormalizable models of this class is again done in two steps. First one replaces the usual interaction of the Boulatov model by tensor invariant interactions to allow for a simple power-counting of the $1/N$ type. Second, one adds to the usual GFT Gaussian measure a Laplacian term to allow for scale analysis. It is however truly non-trivial that renormalization can still work for models of this type. Indeed their propagator $C( g_1, \dots, g_d; g'_1, \dots, g'_d ) $ is def\/initely \emph{not} asymptotic in the ultraviolet regime to a product of delta functions $\delta\big(g_1 (g'_1)^{-1}\big) \cdots \delta\big(g_d (g'_d)^{-1}\big)$ but rather to an average:
\begin{gather*}
C( g_1,\dots, g_d; g'_1, \dots, g'_d ) \simeq_{uv} \int dh \, \delta\big(g_1h (g'_1)^{-1}\big) \cdots \delta\big(g_dh (g'_d)^{-1}\big).
\end{gather*}
Nevertheless an extension of the locality principle still holds for \emph{divergent} graphs, meaning that they can be renormalized again by tensor invariant counterterms. This is possible because the divergent graphs to renormalize, which are melonic graphs, precisely have a suf\/f\/iciently particular structure for such an extended locality property to hold, although it does not hold at all for general graphs.

The set of results include the study of a family of Abelian TGFT models in 4d which was shown to be super-renormalizable for any polynomial interaction~\cite{Carrozza:2012uv}, so is similar in power counting to the family of $P(\phi)_2$ models in ordinary QFT. Abelian just renormalizable models also exist in 5 and 6 dimensions and were studied in~\cite{Samary:2012bw}. They are also asymptotically free \cite{Samary:2013xla}, hence in particular obey the ``quantum relativity principle'' of Section~\ref{qrp} but in which the Hilbert space is the physical one, restricted by the gauge condition. Finally an even more interesting non-Abelian model with group manifold ${\rm SU}(2)$ was successfully renormalized in \cite{Carrozza:2013wda}. For an excellent review of this subject we refer to~\cite{Carrozza:2013mna}, and for an extension to homogenous spaces rather than Lie groups, see~\cite{Lahoche:2015tqa}.

Let us recall that the above studies concern \emph{uncolored} TGFTs. For the colored Boulatov model, Ward--Takahashi identities were derived and interpreted in~\cite{BenGeloun:2011xu}. In~\cite{Geloun:2013zka} the ultraviolet behavior of the same colored theory but with a Laplacian added to the propagator is studied. It is shown that all orders in perturbation theory in the case of the ${\rm U}(1)$ group in three dimensions are convergent; moreover it was convincingly argued that this f\/initeness should also hold for the same model over ${\rm SU}(2)$. Recall however that colored Bosonic models are a priori non-perturbatively unstable, and that colored Fermionic models are therefore more interesting, as they should better behave in that respect. They also have an interesting additional symmetry of rotation between colors~\cite{Gurau:2009tw}.

\section{Asymptotic freedom}\label{asymfree}

This section is essentially reproduced from \cite{Rivasseau:2015ova}. Consider the familiar Laplacian-based normalized Gaussian measure for $d$-dimensional Bosonic f\/ields on ${\rm U}(1)^d$ with periodic boundary conditions
\begin{gather*}
d\mu_C(\phi, \bar \phi) = \left(\prod_{p, \bar p \in \mathbb{Z}^d} \frac{d\phi_p d\bar \phi_{\bar p}}{2i\pi} \right) \operatorname{Det}( C )^{-1}
e^{-\sum_{p, \bar p} \phi_p C^{-1}_{p\bar p}\bar \phi_{\bar p}},
\end{gather*}
where the covariance $C$ is, up to a f\/ield strength renormalization, the inverse of the Laplacian on $S_1^d$ plus a mass term
\begin{gather*}
C_{p,\bar p}=\frac{1}{Z} \frac{\delta_{p,\bar p}}{p^2+ m^2}.
\end{gather*}
Here $p^2 = \sum\limits_{c=1}^d p_c^2 $, $m^2$ is the square of the bare mass, and $Z$ is the so-called wave-function renormalization, which can be absorbed into a $(\phi, \bar \phi) \to \big(Z^{-1/2} \phi, Z^{-1/2} \bar \phi\big)$ f\/ield strength renormalization. If we restrict the indices~$p$, which should be thought as ``momenta", to lie in $[-N, N]^d$ rather than in $\mathbb{Z}^d$ we have proper (f\/inite-dimensional) f\/ields. We can consider $N$ as the ultraviolet cutof\/f, and we are interested in performing the ultraviolet limit $N \to \infty$.

The generating function for the moments of the model is
\begin{gather}
{\mathcal Z}(g,J, \bar J)= \frac{1}{{\mathcal Z}}\int e^{\bar J \cdot \phi + J \cdot \bar \phi } e^{-\frac{g}{2}V (\phi, \bar \phi) } d\mu_C(\phi, \bar \phi) , \label{action}
\end{gather}
where ${\mathcal Z}= {\mathcal Z}(g,J, \bar J)\vert_{J =\bar J =0}$ is the normalization, $g$ is the coupling constant, and the sources $J$ and $\bar J$ are dual respectively to $\bar \phi$ and~$\phi$. The generating function for the connected moments is $W = \log {\mathcal Z}(g,J, \bar J)$.

The simplest quartic vector, matrix and tensor f\/ield theories correspond to this choice of the propagator and only dif\/fer in the \emph{combinatorial} way in which the momenta indices of the four f\/ields branch at the quartic interaction vertex $V (\phi, \bar \phi)$.

The \emph{vector} interaction is the square of the quadratic (mass) term, hence it is \emph{disconnected}, that is it factorizes into two disjoint pairs
\begin{gather}
V_V = \langle \bar \phi , \phi\rangle ^2 =\sum_{p, q} ( \phi_p \bar \phi_{p} )( \phi_q \bar \phi_{q}). \label{quartvect}
\end{gather}
It is just renormalizable for $d=4$.

A \emph{matrix} quartic (connected) interaction is easily def\/ined only for $d=2r$ even. It is obtained by splitting the initial index as a pair $(p,q)$ with $p = (p_1 , \dots, p_r) $, $q= (q_1 , \dots, q_r) $, hence splitting the space ${\mathcal H}_d = {\mathcal H}_r \otimes {\mathcal H}_r$. The f\/ield~$\phi$ is then interpreted as the matrix $\phi_{p q}$, with conjugate matrix $\phi^\star = ^t\bar \phi$ and the vertex $V_M$ is an invariant trace
\begin{gather}
V_M = \operatorname{Tr} \phi \phi^\star \phi \phi^\star = \sum_{p,q, p',q'} \phi_{pq} \bar \phi_{p'q} \phi_{p'q'} \bar \phi_{pq'}. \label{quartmatr}
\end{gather}
It it is just renormalizable for $r=4$, hence $d=8$.

Finally the simplest \emph{tensor} interaction $V_T$ is the color-symmetric sum of \emph{melonic} \cite{Bonzom:2011zz} quartic interactions
\begin{gather}
V_T = \sum_c V_c,\nonumber \\
 V_c(\phi, \bar \phi) = \operatorname{Tr}_c ( \operatorname{Tr}_{\hat c} \phi \bar \phi )^2 =
\sum_{p,\bar p, q,\bar q} \left[ \phi_p\bar \phi_{\bar p} \prod_{c'\neq c}\delta_{p_{c'} \bar p_{c'}} \right] \delta_{p_c \bar q_c} \delta_{q_c \bar p_c}
\left[ \phi_q\bar \phi_{\bar q} \prod_{c'\neq c}\delta_{q_{c'} \bar q_{c'}} \right] , \label{quarttens}
\end{gather}
where $\operatorname{Tr}_{\hat c} \phi \bar \phi$ means partial trace in ${\mathcal H}_d = \otimes_{c=1}^d {\mathcal H}_c$ over all colors except $c$, and $\operatorname{Tr}_c$ means trace over color~$c$. The corresponding model is just renormalizable for $d=5$~\cite{Samary:2014oya}, with a corresponding Connes--Kreimer algebra~\cite{Avohou:2015sia}.

These three dif\/ferent combinatorial models have just renormalizable power counting, like the ordinary scalar $\phi^4_4$ theory. But the class of divergent graphs is more restricted in the combinatorial case. Remark that in all cases the interactions~$V$ are positive for $g >0$. Hence the models are stable for this sign of the coupling constant, which we now assume.

We want to compare the one-loop beta function computation for these just renormalizable quartic vector, matrix and tensor f\/ield theories. This computation is most conveniently performed in the intermediate f\/ield representation.

\begin{figure}[t]\centering
 {\includegraphics[width=0.5\textwidth]{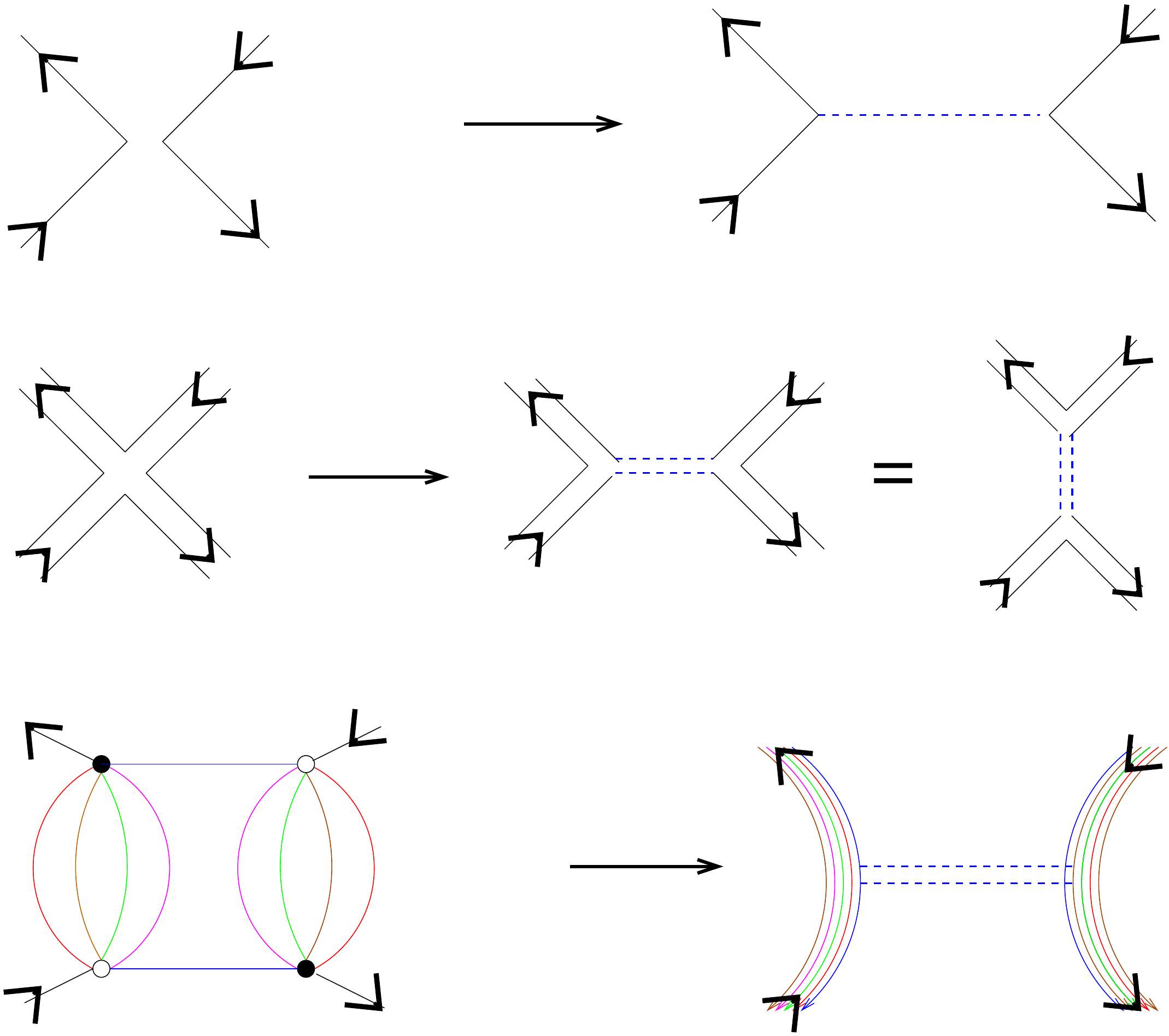}}
 \caption{The vertex is cut in two by the intermediate f\/ield representation. From top to bottom: the vector, matrix and tensor case.
 Incoming and outgoing arrows distinguish $\phi$ and $\bar \phi$.}\label{cutvertex}
\end{figure}

An intermediate f\/ield $\sigma$ splits the quartic vertex in two halves, as pictured in Fig.~\ref{cutvertex}, through the simple integral representation
\begin{gather}
 e^{-\frac{g}{2}V (\phi, \bar \phi) } = \int d\nu(\sigma) e^{i \sqrt g \bar \phi\phi \cdot \sigma}. \label{interm1}
\end{gather}
In this formula $d\nu$ is a Gaussian measure with covariance~1 on the intermediate f\/ield $\sigma$. The three cases \eqref{quartvect}--\eqref{quarttens} lead to $\sigma$ f\/ields of dif\/ferent nature and to dif\/ferent combinatorial rules for the dot in~\eqref{interm1}. In the vector case the $\sigma$ f\/ield is a \emph{scalar}, ref\/lecting the already factorized nature of~\eqref{quartvect}. In the matrix and tensor case~\eqref{quartmatr},~\eqref{quarttens} $\sigma$ is a \emph{matrix}. More precisely, in the matrix case~\eqref{quartmatr}, it is a single matrix with its two arguments in ${\mathbb Z}^4$. In the tensor case~\eqref{quarttens}, it is the sum of f\/ive dif\/ferent colored matrices $\sigma_c$ with their two arguments in~${\mathbb Z}$, one for each color~$c$, and should be properly written as
\begin{gather*} {\sigma } = \sum_c \sigma_c \otimes \mathbb{I}_{\hat c} .
\end{gather*}

The advantage of this representation is that the $\bar\phi$ and $\phi$ functional integral becomes quadratic, hence can be performed explicitly, yielding
\begin{gather*}
{\mathcal Z}(g,J, \bar J) = \frac{1}{{\mathcal Z}} \int d\nu(\sigma) \int d\mu_{C}(\phi, \bar \phi) e^{i \sqrt g \bar \phi\phi \cdot \sigma}
e^{\bar J \cdot \phi + J \cdot \bar \phi } \nonumber \\
\hphantom{{\mathcal Z}(g,J, \bar J)}{} = \frac{1}{{\mathcal Z}} \int d\nu(\sigma) e^{\langle \bar J , C^{1/2} R (\sigma) C^{1/2} J \rangle - \operatorname{Tr} \log \left[\mathbb{I}-i \sqrt g C^{1/2} {\sigma } C^{1/2} \right] } ,
\end{gather*}
where $R$ is the symmetric \emph{resolvent} operator
\begin{gather*}
R (\sigma) \equiv \frac{1}{ \mathbb{I} -i \sqrt g C^{1/2} \sigma C^{1/2} } .
\end{gather*}

In all cases the one-loop beta function boils down to the same computation, up to subtle dif\/ferences of a purely combinatorial nature. Let us call~$\Gamma_{2p}$ the $2p$-point vertex function, hence the sum of one-particle irreducible amputated Feynman amplitudes with $2p$ external legs. The renormalized BPHZ prescriptions are def\/ined by momentum space subtractions at zero momentum, which we can restrict to \emph{divergent} graphs (e.g., planar in the matrix case, melonic in the tensor case)
\begin{gather}
\frac{g_r}{2} = - \Gamma_{4}(0),\qquad Z-1 = \left[ \frac{\partial}{\partial p^2} \Gamma_2 \right] (0), \label{gammas}
\end{gather}
where $g_r$ is the renormalized coupling. Performing the f\/ield strength renormalization we can rescale to 1 the wave function renormalization at high ultraviolet cutof\/f at the cost of using a \emph{rescaled} bare coupling $g'_b = Z^{-2} g_b$. The one-loop $\beta_2$ coef\/f\/icient shows how this rescaled bare coupling evolves at f\/ixed~$g_r$ when $N \to \infty$. It writes
\begin{gather} g'_b = g_r \big[1 + \beta_2 g_r ( \log N + {\rm f\/inite}) + O\big(g_r^2\big) \big], \label{beta}
\end{gather}
where $N$ is the ultraviolet cutof\/f, and ``f\/inite'' means bounded as $N \to \infty$. As well-known $\beta_2 > 0$ corresponds to a coupling constant which f\/lows out of the perturbative regime in the ultraviolet. $\beta_2 < 0$ corresponds to asymptotic freedom: the (rescaled) bare coupling f\/lows to zero as $N \to \infty $. $\beta_2 =0$ is inconclusive as the analysis of the renormalization group f\/low must be pushed further, but, if reproduced at higher orders, indicates a fully scale invariant theory.

It is easier to compute the bare perturbation theory, as it does not involve any subtraction. Starting from \eqref{gammas}, we f\/ind that $\Gamma_{4}$ and $Z-1$ always involve the \emph{same} logarithmically divergent sum, namely
\begin{gather}
\sum_{q\in[-N,N]^4}\frac{1}{\big(q^2+m_r^2\big)^2} = 2 \pi^2 \log N + {\rm f\/inite} , \label{beta3}
\end{gather}
where $m_r^2= Zm^2 -\Gamma_{2} (0)$ is the renormalized mass.
However this sum arises with various combinatoric coef\/f\/icients. More precisely
\begin{gather}
\Gamma_{4} (0) = -\frac{g_b}{2} \left[1 - a g_b \sum_{q\in[-N,N]^4}\frac{1}{\big(q^2+m_r^2\big)^2}+O\big(g_b^2\big)\right] , \label{beta1}
\\
Z = 1 + \frac{\partial \Gamma_{2}}{\partial p^2} \Big|_{p=0} = 1 + b g_b \sum_{q\in[-N,N]^4}\frac{1}{\big(q^2+m_r^2\big)^2}+O\big(g_b^2\big), \label{beta2}
\end{gather}
where $a$ and $b$ are combinatoric coef\/f\/icients that depend on the particular case (vector, matrix or tensor) considered.

Since $g'_b = Z^{-2} g_b$, multiplying \eqref{beta} by $Z^2$ and taking into account \eqref{gammas}--\eqref{beta2}, which imply $g_r = g_b + O(g_b^2)$ and $Z= 1 + O(g_b)$, we f\/ind
\begin{gather*}
 Z^2\Gamma_4 (0) = -\frac{g_b}{2} \big[1 - \beta_2 g_b ( \log N + {\rm f\/inite}) + O\big(g_b^2\big)\big],
\end{gather*}
hence in all cases
\begin{gather*} \beta_2 = (a-2b) 2 \pi^2 .
\end{gather*}
We are left with the simple problem of computing the coef\/f\/icients $a$ and $b$ of the one loop leading diagrams for~$\Gamma_4$ and~$Z$.

\begin{figure}[t]\centering
 \includegraphics[width=0.6\textwidth]{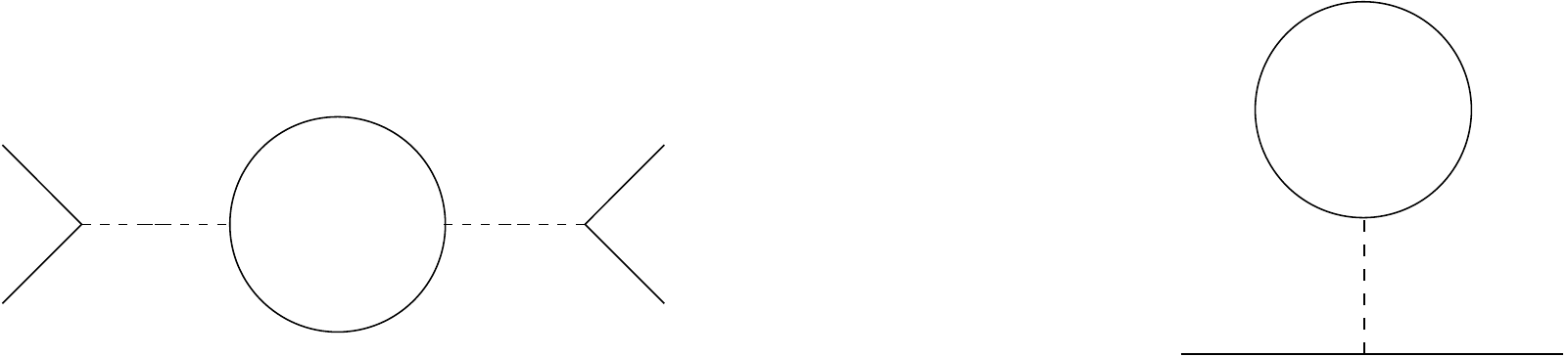}
\caption{The (single) one loop melonic graph in the tensor case for $\Gamma_4$ and $\Gamma_2$ are trees for the intermediate f\/ield (dashed) lines.} \label{onelooptensor}
\end{figure}

\begin{figure}[t]\centering
 \includegraphics[width=0.65\textwidth]{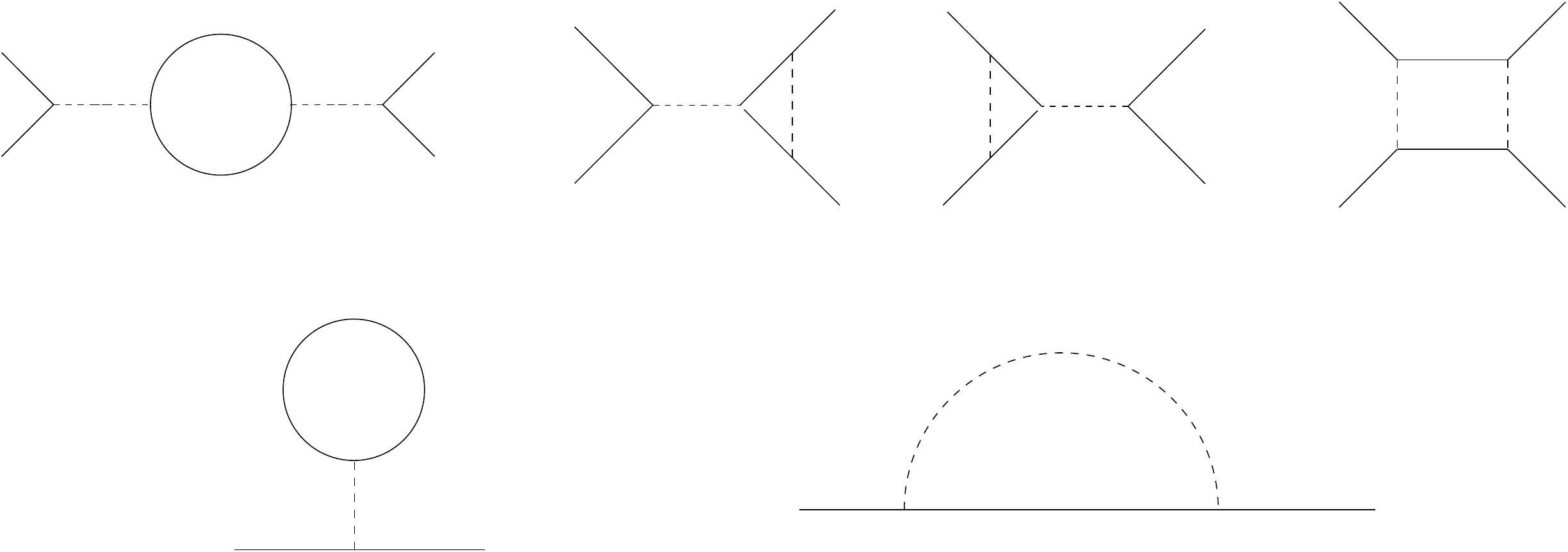}
 \caption{The dominant one loop graphs in the matrix case for $\Gamma_4$ and $\Gamma_2$ are planar in the intermediate f\/ield representation.}
 \label{oneloopmatrix}
\end{figure}

\begin{itemize}\itemsep=0pt
\item
In the vector case, $a=1$. The only divergent graph is the one on the left of Fig.~\ref{onelooptensor}. Resolvents (derivatives with respect to~$J$ and $\bar J$) come up with a factor 1 and no sym\-met\-ry factorial, whether terms from the $\operatorname{Tr} \log$ expansion come up with a factor~$1/n$ for a~$ \operatorname{Tr} \big( i \sqrt g C^{1/2} {\vec \sigma }C^{1/2}\big)^n $ (because of the Taylor series of the logarithm), plus a symmetry factor~$1/k!$ if there are~$k$ of them (this factor comes from expansion of the exponential). The combinatoric weight for the tree graph at order~$ g^2$ for~$\Gamma_4$ is therefore~$1$, which decomposes into a $1/2$ for the single loop vertex ($n=2$, $k=1$) times a factor~2 for the two Wick contractions.
In this vector case $b=0$ since the one loop tadpole on the right of Fig.~\ref{onelooptensor}, the only contributing graph at order~$g$, does not have any external momentum dependence. Hence
\begin{gather*}
 \left[ \frac{\partial}{\partial p^2} \Gamma_2 \right] (0) = O\big(g^2\big)\quad \Rightarrow \quad b=0.
\end{gather*}
Hence $\beta_2 = 2 \pi^2 $ and the theory has no UV f\/ixed point, at least in this approximation.

\item In the tensor case, $a=1$ for the same reasons than in the vector case. Indeed for any of the f\/ive melonic interactions there is a single divergent graph of the corresponding color of the type pictured on the left of Fig.~\ref{onelooptensor}.
But now we also have $b=1$. Indeed remark f\/irst that $b>0$ because two minus signs compensate, one in front of $g$ in~\eqref{action} and the other coming from the mass subtraction, since
\begin{gather*}
\left[\frac{1}{p_c^2} \left( \frac{1}{q^2 + p_c^2 + m_r^2} - \frac{1}{q^2 + m_r^2} \right) \right]_{p_c = 0}
= - \frac{1}{ \big( q^2 + m_r^2 \big)^2} .
\end{gather*}
The combinatorics is then 1 because there is a single loop vertex with $n=1$, $k=1$ and a single Wick contraction to branch it on the external resolvent as shown for the graph on the right of Fig.~2. Summing over the colors $c$ of this contraction simply reconstructs $\sum_c p_c^2 =p^2$. In conclusion $\beta_2 = -2 \pi^2 $ and the theory is asymptotically free, in agreement with~\cite{BenGeloun:2012pu,Benedetti:2014qsa,Samary:2014oya}. Wave function, or f\/ield strength renormalization won over coupling constant renormalization because of the square power in~$Z^2$.

\item In the matrix case, the vertex crossing symmetry means more terms diverge logarithmically than in the vector and tensor cases, namely those corresponding to planar \emph{maps} in the intermediate representation. The crossing symmetry is a~${\mathbb Z}_2$ symmetry, but it acts dif\/ferently on~$\Gamma_4$ and~$\Gamma_2$. Since the one loop graph for~$\Gamma_4$ has \emph{two} vertices, hence two (dotted) $\sigma$ propagators, the crossing symmetry acts twice independently and generates an orbit of four planar maps, represented in the top part of Fig.~\ref{oneloopmatrix}. In contrast the crossing symmetry acts only once on the orbit of the $\Gamma_2$ term, generating only the two planar maps pictured in the bottom part of Fig.~\ref{oneloopmatrix}. Hence $a=4$ and $b=2$ which leads to $\beta_2 = 0$! This combinatorial ``miracle'' persists at all orders: in fact the logarithmically divergent part of $ Z^2 \Gamma_4 (0)$ is exactly~0 at all orders in~$g$, as can be shown through combining a Ward identity with the Schwinger--Dyson equations of the theory~\cite{Disertori:2006nq}. The corresponding theory is asymptotically \emph{safe}.
\end{itemize}

\section{Conclusion}

In conclusion we would like to recall the main positive mathematical aspects of the tensor track:
\begin{itemize}\itemsep=0pt
\item it proposes to perform a sum both over any topology and any smooth structure up to dimension/rank~4, following the paradigm of Feynman functional integral quantization (``sum over histories''),

\item this sum is naturally pondered by a discrete version of the Einstein--Hilbert action,

\item it is possible to introduce an abstract notion of scales through Laplacian-type operators,

\item the tensor symmetry is compatible with renormalization and acts as a substitute for locality,

\item it supports a power counting tool, namely the $1/N$ tensor expansion,

\item renormalization group f\/lows can be investigated in the tensor theory space,

\item the simplest models are asymptotically free, hence obey therefore to the natural extension of the general relativity principle sketched above: physics in the extreme ultraviolet limit becomes asymptotically independent of any preferred basis in the huge Hilbert space of states of the universe,

\item constructive control is possible at least in simple (super-renormalizable) cases.

\end{itemize}

Let us immediately temper however all these positive points with a lot of caveats. Of course enormous work lies ahead of the tensor track program, which is only one among many compe\-ting approaches to quantum gravity. Among the major problems to tackle in our approach we can list f\/inding Euclidean axioms including the right generalization of the Osterwalder--Schrader positivity axiom, to allow in particular emergence of Lorentzian time and causality; constructive treatment of more than quartic interactions; renormalization group evolution from the arborescent to more realistic macroscopic phases (see~\cite{Bonzom:2015axa} for a step in that direction); consequences of the theory for cosmology scenarios and for black holes; and addition of the standard model matter f\/ields to the picture.

Altogether we nevertheless have the impression that the tensor track has matured enough to be taken seriously and explored further. At the physical level it suggests an emergent space-time scenario with an initial arborescent phase of the universe. This result should not be immediately discarded as non-physical, since the richness of subdominant tensor interactions could lead this arborescent phase to evolve later into geometries closer to our actual universe.

\subsection*{Acknowledgements}
We thank R.~Avohou, D.~Benedetti, J.~Ben Geloun, V.~Bonzom, S.~Carrozza, S.~Dartois, T.~Delepouve, O.~Samary Dine, R.~Gurau, T.~Krajewski,
V.~Lahoche, L.~Lionni, D.~Oriti, A.~Tanasa, F.~Vignes-Tourneret and R.~Wulkenhaar for discussions and contributions on many aspects of the tensor track program.

\pdfbookmark[1]{References}{ref}
\LastPageEnding

\end{document}